\documentclass[journal=jacsat,manuscript=article]{achemso}
\usepackage{chemformula} 
\usepackage[T1]{fontenc} 
\usepackage{graphicx}%
\usepackage{multirow}%
\usepackage{amsmath,amssymb,amsfonts}%
\usepackage{amsthm}%
\usepackage{mathrsfs}%
\usepackage[title]{appendix}%
\usepackage{xcolor}%
\usepackage{textcomp}%
\usepackage{manyfoot}%
\usepackage{booktabs}%
\usepackage{algorithm}%
\usepackage{algorithmicx}%
\usepackage{algpseudocode}%
\usepackage{listings}%

\usepackage[english]{babel}
\usepackage[T1]{fontenc}
\usepackage[latin9]{inputenc}
\usepackage{amsmath,amssymb,graphicx,mathrsfs,xcolor,babel,textcomp,esint}

\newcommand{\tmop}[1]{\ensuremath{\operatorname{#1}}}

\author{Ziyang Gan}
\affiliation{Shanghai Advanced Research Institute, Chinese Academy of Sciences, Shanghai 201210, China}
\affiliation{Center for Transformative Science, School of Physical Science and Technology, ShanghaiTech University, Shanghai 201210, China}
\affiliation{University of Chinese Academy of Sciences, Beijing 100049, China}

\author{Kaixuan Zhang}
\affiliation{Zhejiang Provincial Key Laboratory of Ultra-Weak Magnetic-Field Space and Applied Technology, Hangzhou Innovation Institute, Beihang University, Hangzhou 310051, China}
\affiliation{Shanghai Advanced Research Institute, Chinese Academy of Sciences, Shanghai 201210, China}

\author{Yuan Gao}
\affiliation{Shanghai Advanced Research Institute, Chinese Academy of Sciences, Shanghai 201210, China}

\author{Ahai Chen}
\affiliation{Center for Transformative Science, School of Physical Science and Technology, ShanghaiTech University, Shanghai 201210, China}

\author{Yizhu Zhang}
\email{zhangyizhu@tju.edu.cn}
\affiliation{Center for Terahertz Waves and College of Precision Instrument and Optoelectronics Engineering, Key Laboratory of Opto-electronics Information and Technical Science, Ministry of Education, Tianjin University, Tianjin 300350, China}
\affiliation{Shanghai Advanced Research Institute, Chinese Academy of Sciences, Shanghai 201210, China}
\affiliation{Max-Planck Institute for Nuclear Physics, Heidelberg 69117, Germany}
\author{Tian-Min Yan}
\email{yantm@sari.ac.cn}
\affiliation{Shanghai Advanced Research Institute, Chinese Academy of Sciences, Shanghai 201210, China}
\author{Thomas Pfeifer}
\affiliation{Max-Planck Institute for Nuclear Physics, Heidelberg 69117, Germany}
\author{Yuhai Jiang}
\email{jiangyh3@shanghaitech.edu.cn}
\affiliation{Center for Transformative Science, School of Physical Science and Technology, ShanghaiTech University, Shanghai 201210, China}
\affiliation{Shanghai Advanced Research Institute, Chinese Academy of Sciences, Shanghai 201210, China}

\title[An \textsf{achemso} demo]
  {Probing Electronic Motion and Core Potential by Coulomb-reshaped Terahertz Radiation}

\abbreviations{IR,NMR,UV}
\keywords{American Chemical Society, \LaTeX}

\begin{document}

\begin{tocentry}

\begin{figure}[H]
\includegraphics[width=0.990\textwidth]{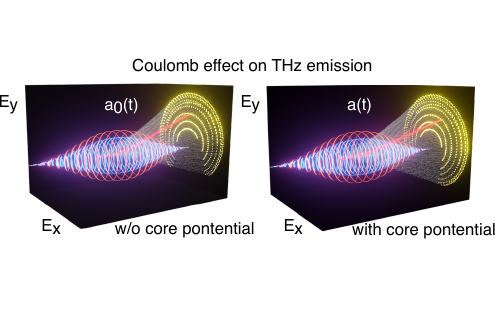}
\end{figure}


\end{tocentry}

\begin{abstract}
The nature of electronic motion and structural information of atoms and
molecules is encoded into strong-field induced radiations ranging from terahertz
(THz) to extreme ultraviolet wavelength. 
The dependence of THz yields in bi-chromatic laser fields on ellipticity and interpulse phase delay were experimentally measured, and the trajectory calculations establish the link between the THz emission and the motion of the photoelectron wave packet. 
The interaction between the photoelectron and parent core transforms from soft collision to recollision as the laser field tuned from elliptical to linear polarization, which can be reflected in THz emission.
The soft collision is found to be more effective in reconstructing electron dynamics through THz polarization, which enables to construct the effective core potential of the generating medium with the Coulomb-reshaped THz radiation in an elliptically polarized laser field. 
Our work allows designing innovative all-optical THz measurements of electronic and structural dynamics.
\end{abstract}

\section{Introduction}\label{sec:introduction}

Strong-field induced radiation ranging from terahertz (THz) to extreme ultraviolet wavelength contain a wealth of structural and dynamical information of the generating medium. The high-harmonic generation (HHG) of extreme ultraviolet photons has been widely used for molecular orbital tomography \cite{itatani_tomographic_2004,vozzi_generalized_2011}, and the probing of electron wave packet \cite{shafir_resolving_2012,pedatzur_attosecond_2015},  nuclear dynamics and structural rearrangement on a subfemtosecond time scale \cite{lein_attosecond_2005,baker_probing_2006,bian_probing_2014}. Analogous to HHG, the THz wave generation (TWG) \cite{cook_intense_2000},  so-called zeroth-order Brunel harmonics \cite{Brunel_Harmonic_1990}, has been also considered as an all-optical approach for probing molecular structures \cite{huang_joint_2015}, and recently an innovative optical attoclock \cite{babushkin_all-optical_2022} where the THz polarization direction acts as a "clock hand" for mapping the tunneling delay, laterally complementing to currently used attoclock implemented by photoelectron momentum spectroscopy \cite{Eckle_attosecond_2008a,Eckle_attosecond_2008b,camus_experimental_2017,Sainadh_attosecond_2019}.

The strong-field induced TWG physically originates from the acceleration of tunneling photoelectron wave packet in oscillating electric field, described macroscopically by the photo-current (PC) model \cite{kim_terahertz_2007}, or microscopically by continuum-continuum transition in strong field approximation \cite{zhou_terahertz_2009,zhang_continuum_2020}.  In typical scenarios, when fitting the macroscopic THz yield, the aforementioned models neglecting the Coulomb interaction between the free electron and the parent core suffices. Nevertheless, when employing all-optical THz probing to investigate microscopic structures and electron dynamics, the Coulomb effects becomes highly sensitive and thus necessitates careful consideration.

The Coulomb influence on 
photoelectron momentum spectra and HHGs, when retrieving the photoelectron dynamics, has been fully noticeable. When measuring the electron tunneling delay implemented by attoclock \cite{Eckle_attosecond_2008b,camus_experimental_2017,Sainadh_attosecond_2019}, all the works emphasized that, the Coulomb interaction must be taken into account, and the tunneling delay has to be correctly disentangled from the final photoelectron momentum spectra to achieve a meaningful quantitative interpretation.
The Coulomb-reshaped electronic wave packet has been encoded in the phase of HHGs \cite{Azoury_electronic_2019}, which affects the accuracy of structural tomography \cite{Boutu_2008,Zhou_2008,vozzi_generalized_2011}.  For TWGs, the involvement of Coulomb potential in microscopic information extraction, as well as the speculation on photoelectron motion and structural information, has been rarely investigated.

Although the TWG has been well described by the PC model
\cite{kim_terahertz_2007,kim_coherent_2008}, where the net residual
photocurrent density plays the core role, the photocurrent lacks more
fine-grained microscopic information of photoelectron dynamics. As the
photocurrent is essentially a macroscopic correspondence of asymmetric photoelectron wave packet conceptually described by an ensemble of propagating photoelectron trajectories \cite{Babushkin_2011,liu_trajectory-based_2014}, the TWG can be
evaluated from microscopic trajectories to account for the influence from both
the external field $\boldsymbol{E} (t)$ and the potential of the parent ion $V
(\boldsymbol{r})$ \cite{zhang_synchronizing_2012}. 
This influence is experimentally substantiated by measuring the optimal THz yields as a function of two-color phase delay \cite{zhang_synchronizing_2012}, as well as by examining the THz polarization under specific polarization combinations of two-color fields \cite{gao_2023_coulomb}. 

In this work, according to the classical trajectory Monte-Carlo (CTMC) method,  we elucidate how various types of electron-core interactions, including soft collisions and re-collisions, are imprinted in the TWG polarization. Meanwhile, it reveals the equivalence between the THz polarization and asymmetry pointer of photoelectron momentum distributions (PMDs), which can be substantially employed for the reconstruction of atomic core potentials.

Strong-field induced radiation ranging from terahertz (THz) to extreme ultraviolet wavelength contain a wealth of structural and dynamical information of the generating medium. The high-harmonic generation (HHG) of extreme ultraviolet photons has been widely used for molecular orbital tomography \cite{itatani_tomographic_2004,vozzi_generalized_2011}, and the probing of electron wave packet \cite{shafir_resolving_2012,pedatzur_attosecond_2015},  nuclear dynamics and structural rearrangement on a subfemtosecond time scale \cite{lein_attosecond_2005,baker_probing_2006,bian_probing_2014}. Analogous to HHG, the THz wave generation (TWG) \cite{cook_intense_2000},  so-called zeroth-order Brunel harmonics \cite{Brunel_Harmonic_1990}, has been also considered as an all-optical approach for probing molecular structures \cite{huang_joint_2015}, and recently an innovative optical attoclock \cite{babushkin_all-optical_2022} where the THz polarization direction acts as a "clock hand" for mapping the tunneling delay, laterally complementing to currently used attoclock implemented by photoelectron momentum spectroscopy \cite{Eckle_attosecond_2008a,Eckle_attosecond_2008b,camus_experimental_2017,Sainadh_attosecond_2019}.

The strong-field induced TWG physically originates from the acceleration of tunneling photoelectron wave packet in oscillating electric field, described macroscopically by the photo-current (PC) model \cite{kim_terahertz_2007}, or microscopically by continuum-continuum transition in strong field approximation \cite{zhou_terahertz_2009,zhang_continuum_2020}.  In typical scenarios, when fitting the macroscopic THz yield, the aforementioned models neglecting the Coulomb interaction between the free electron and the parent core suffices. Nevertheless, when employing all-optical THz probing to investigate microscopic structures and electron dynamics, the Coulomb effects becomes highly sensitive and thus necessitates careful consideration.

The Coulomb influence on 
photoelectron momentum spectra and HHGs, when retrieving the photoelectron dynamics, has been fully noticeable. When measuring the electron tunneling delay implemented by attoclock \cite{Eckle_attosecond_2008b,camus_experimental_2017,Sainadh_attosecond_2019}, all the works emphasized that, the Coulomb interaction must be taken into account, and the tunneling delay has to be correctly disentangled from the final photoelectron momentum spectra to achieve a meaningful quantitative interpretation.
The Coulomb-reshaped electronic wave packet has been encoded in the phase of HHGs \cite{Azoury_electronic_2019}, which affects the accuracy of structural tomography \cite{Boutu_2008,Zhou_2008,vozzi_generalized_2011}.  For TWGs, the involvement of Coulomb potential in microscopic information extraction, as well as the speculation on photoelectron motion and structural information, has been rarely investigated.

Although the TWG has been well described by the PC model
\cite{kim_terahertz_2007,kim_coherent_2008}, where the net residual
photocurrent density plays the core role, the photocurrent lacks more
fine-grained microscopic information of photoelectron dynamics. As the
photocurrent is essentially a macroscopic correspondence of asymmetric photoelectron wave packet conceptually described by an ensemble of propagating photoelectron trajectories \cite{Babushkin_2011,liu_trajectory-based_2014}, the TWG can be
evaluated from microscopic trajectories to account for the influence from both
the external field $\boldsymbol{E} (t)$ and the potential of the parent ion $V
(\boldsymbol{r})$ \cite{zhang_synchronizing_2012}. 
This influence is experimentally substantiated by measuring the optimal THz yields as a function of two-color phase delay \cite{zhang_synchronizing_2012}, as well as by examining the THz polarization under specific polarization combinations of two-color fields \cite{gao_2023_coulomb}. 

In this work, according to the classical trajectory Monte-Carlo (CTMC) method,  we elucidate how various types of electron-core interactions, including soft collisions and re-collisions, are imprinted in the TWG polarization. Meanwhile, it reveals the equivalence between the THz polarization and asymmetry pointer of photoelectron momentum distributions (PMDs), which can be substantially employed for the reconstruction of atomic core potentials.

\section{Coulomb effect on THz wave emission}\label{sec:core-potential-effect}

In CTMC method, within an ensemble of
trajectories $\{i\}$, the $i$th trajectory $\boldsymbol{r}_i (t)$ is determined
by solving the equation of motion $\boldsymbol{a}_i (t) \equiv \partial^2
\boldsymbol{r}_i (t) / \partial t^2 = -\boldsymbol{E} (t) - \nabla V
[\boldsymbol{r}_i (t)]$ starting from the initial tunneling time $t_0^{(i)}$. 
The initial conditions and ionization rate $w_i$ of the CTMC method are
detailed in \emph{Supplement} Sec. 2. 
The radiation derives from the
acceleration of the ensemble $\boldsymbol{a} (t) = \sum_i w_i \Theta (t -
t_0^{(i)}) \boldsymbol{a}_i (t)$ with 
$\Theta (t)$ the Heaviside function  \cite{zhang_electron_2021,Fan_2023}. The time-domain THz wave is
obtained by evaluating $\mathscr{F}^{- 1} \{ \mathscr{W}  \mathscr{F}
\{\boldsymbol{a}(t)\}(\omega)\} (t)$ with $\mathscr{F}$ the Fourier transform
and $\mathscr{W}$ the low-pass filter.

\begin{figure}[h]
\includegraphics[width=1\textwidth]{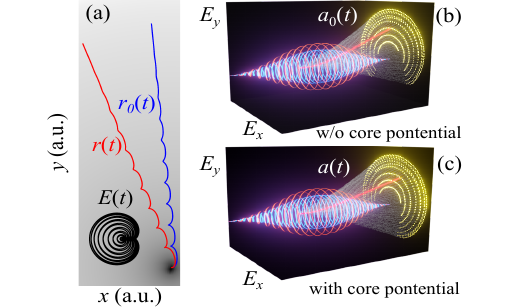}
\caption{The influence of Coulomb interaction on photoelectron trajectories. 
Panel (a) shows an exemplary trajectory of photoelectron, either subjected to the Coulomb potential (red) or not (blue).  
Panels (b) and (c) show the distribution of photoelectron wave packet as an ensemble of classical trajectories with and without Coulomb interaction as indicated.
}\label{fig:wavepacket}
\end{figure}

Taking the two-color bi-circularly
polarized fields for instance, once the electron is released from the atom,
its trajectory, compared to the path ${\bf r}_0(t)$ driven solely by the external light field, may become a bent one ${\bf r}(t)$ in the presence of Coulomb interaction, as shown in Fig. \ref{fig:wavepacket}(a).
The Coulomb effects on each trajectory eventually alters
the global distribution of the electronic wave packet. Fig. \ref{fig:wavepacket}(b)
shows the ensemble of trajectories when the Coulomb potential is absent. At an
arbitrary time $t$, the positions of all classical trajectories, 
$r^{(i)}_0 (t)$, correspond to the spatial distribution of photoelectron
wave packet. The radiation is induced by the ensemble acceleration $\langle \boldsymbol{a}_0
(t) \rangle = -\boldsymbol{E} (t) n (t)$, showing a consistent form to
the PC model \cite{kim_terahertz_2007}, but with the electron density $n (t)
= \sum_i w_i \Theta (t - t_0^{(i)})$ as a sum over all trajectories. When the
parent ion is present, the distribution of trajectories is slightly distorted
as shown in Fig. \ref{fig:wavepacket}(c). The distortion may result in
observable patterns in PMD, and also equips
the acceleration with a correction term, $\langle \boldsymbol{a}(t) \rangle =
\langle \boldsymbol{a}_0 (t) \rangle + \langle \boldsymbol{a}_C (t) \rangle$, with
$\langle \boldsymbol{a}_C (t) \rangle = \sum_i w_i \Theta (t - t_0^{(i)})  \frac{\boldsymbol{r}_i
(t)}{|\boldsymbol{r}^3_i (t) |}$ from the Coulomb potential, inducing extra
modulation in radiation.

\begin{figure}[h]
  \includegraphics[width=1\textwidth]{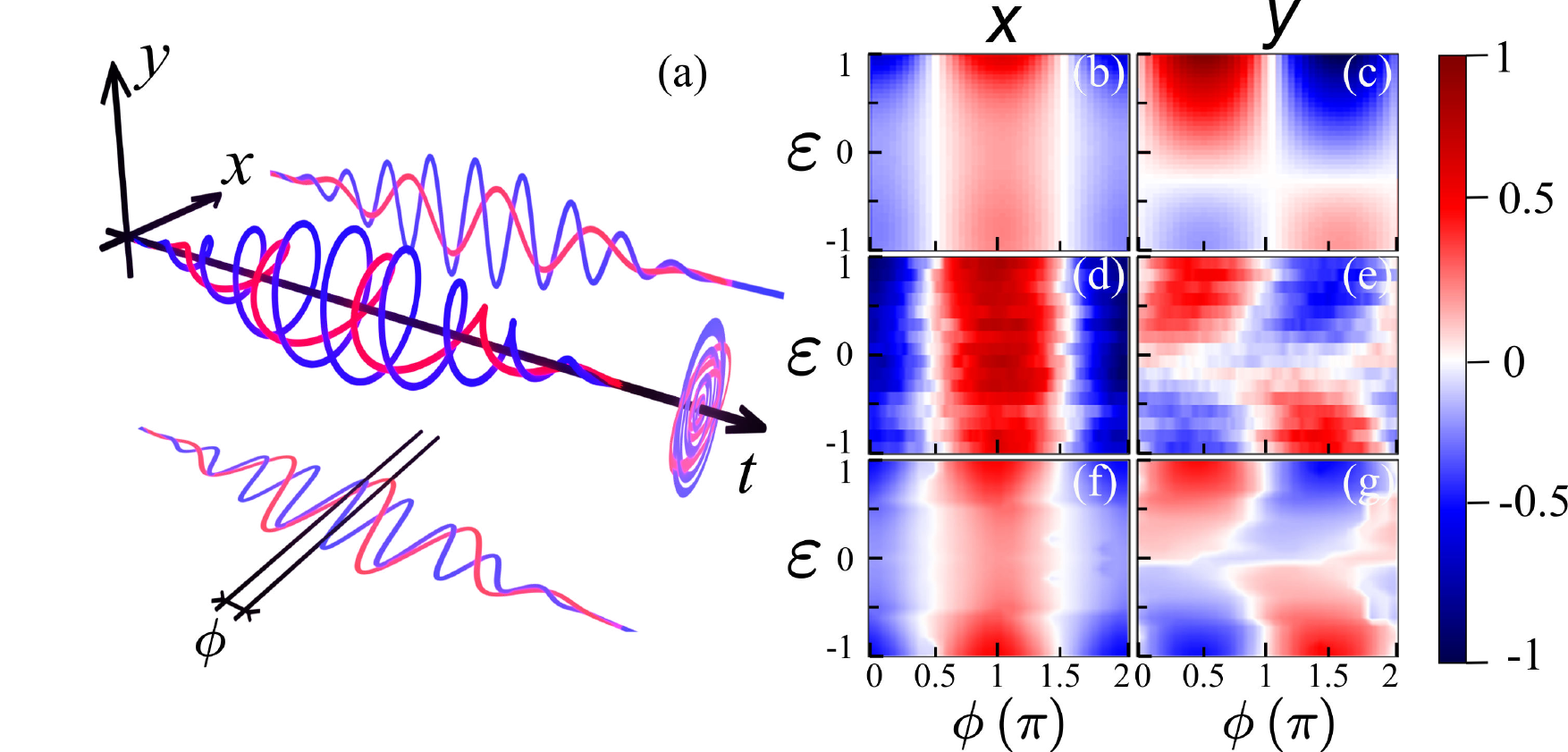}
  \caption{The THz peak-to-peak distributions in
  two-color fields $S_{\text{pp},\sigma} (\varepsilon, \phi)$. 
  Panel (a) presents an illustration of the laser fields, 
  where the $\omega$-field (red) is elliptically polarized with a ellipticity $\varepsilon$ and the $2 \omega$-field (blue) is circularly polarized. 
  The $\omega$-$2 \omega$ phase delay is $\phi$. The
  $x$ direction is defined as parallel to the polarization of the
  $\omega$-field when $\varepsilon = 0$ (see \emph{Supplement} Sec. 1 for
  detailed definition). The PP distributions in the $x$ direction,
  $S_{\text{pp}, x} (\varepsilon, \phi)$, are shown for (b) PC model, (d)
  experiment, and (f) CTMC. Correspondingly, the PP distributions in the $y$ direction, $S_{\text{pp}, y} (\varepsilon, \phi)$, are shown in Panels (c), (e), and (g).}\label{fig:PP-distributions}
\end{figure}

We explore TWG in two-color fields by mixing the fundamental
of a Ti:sapphire laser [800 nm ($\omega$), 35 fs] with its second harmonic [400 nm ($2 \omega$), circularly polarized].
The $\omega$ and $2\omega$ beams have intensities of $I$=1.5$\times 10^{14}$ W/cm$^2$ and $I/2$, respectively.
As schematically demonstrated
in Fig. \ref{fig:PP-distributions}(a), we measure the THz yield $S
(\varepsilon, \phi)$ as a function of $\varepsilon$, the ellipticity of
$\omega$ beam, and $\phi$, the interpulse phase delay. 
The TWGs are detected with electro-optic sampling, and the polarization components of time-domain waveform $E_{\text{THz}, \sigma} (t)$ ($\sigma$$\equiv$$x,y$)
are recorded. 
Defining the THz
peak-to-peak (PP) amplitude, $S_{\text{pp}, \sigma}$=$\pm | \mathrm{max}
[E_{\text{THz}, \sigma} (t)] - \mathrm{min} [E_{\text{THz}, \sigma} (t)] |$,
we measure the dependence of $S_{\text{pp}, \sigma}$ on $\varepsilon$$\in$[0, 1] 
and $\phi$$\in$[0, 2$\pi$]. 
As the absolute time zero of $\phi$ is technically challenging to determine in the experiment, we further propose a method based on CTMC to determine $\phi$ by measuring TWG polarization directions (see subsequent parts).
The detailed experimental setup, raw data and self-referencing method are presented in \emph{Supplement} Sec. 1.

Fig. \ref{fig:PP-distributions}(b)-(g) present the distributions of $S_{\text{pp},
\sigma} (\varepsilon, \phi)$ obtained from the PC model, experiment, and CTMC method. 
Contrary to the distribution of
$S_{\text{pp}, y}  (\varepsilon, \phi)$ evaluated by the PC model
shown in
Fig. \ref{fig:PP-distributions}(c), the experimental result in (e) 
exhibits 
a bend along $\varepsilon$, which can be 
replicated 
by the CTMC calculation in
Panel (g). 
The CTMC calculations without Coulomb potential show same results
as the PC model, confirming the equivalence between the two methods.
Thus, the bend
by comparison shown in Fig. \ref{fig:PP-distributions} is confirmed to be attributed to the Coulomb effects.

The bend of $S_{\text{pp,$y$}}$ induced by Coulomb potential is more pronounced than that of $S_{\text{pp,$x$}}$, because, along the $y$ direction, the contribution of the Coulomb potential to the photoelectron momentum is comparable to the momentum induced by the laser field.

Through CTMC, the TWG is closely tied to the photoelectron wave packet as an ensemble of trajectories. The CTMC model establishes a correlation between the TWGs and PMDs. The THz emission ($\boldsymbol{E}_{\text{THz}}$) from the trajectory ensemble is expressed as the summation of individual trajectories as 
\begin{eqnarray}
\boldsymbol{E}_{\text{THz}} (\omega \rightarrow 0) 
&=& \sum_{i} {w_{i} \bigg( \lim_{\omega \rightarrow 0}  \int_{-\infty}^{\infty} \mathrm{d} t \frac{\partial \boldsymbol{v}_{i} (t)}{\partial t} \mathrm{e}^{- \mathrm{i} \omega t} \bigg) } \nonumber \\
&=& \sum_{i} {w_{i}  \boldsymbol{v}_{i} (\infty)}.
\end{eqnarray}
Here, $\omega$ represents the frequency of strong-field induced radiation, $\boldsymbol{v}_{i} (t)$ and $\boldsymbol{v}_{i} (\infty)$ denote the instantaneous velocity and asymptotic velocity (drift velocity) of the $i$th trajectory. The contribution of the $i$th trajectory to TWG is expressed as $\boldsymbol{E}_{\text{THz},i} \propto w_{i} \boldsymbol{v}_{i} (\infty)$. The polarization direction of $\boldsymbol{E}_{\text{THz},i}$ is defined as $\boldsymbol{\hat{E}}_{\text{THz},i}$, equivalent to the direction of asymptotic velocity $\boldsymbol{\hat{v}}_{i} (\infty)$. The amplitude of $\boldsymbol{E}_{\text{THz},i}$ is defined as $|\boldsymbol{E}_{\text{THz},i}| \propto |w_{i} \boldsymbol{v}_{i} (\infty)|$. 
We define the asymmetry of PMDs as ${\bf P}_e = \sum_{i} w_{i}  \boldsymbol{v}_{i} (\infty)$ with an asymmetrical direction of ${\bf \hat{P}}_e$. 
$\boldsymbol{E}_{\text{THz}}$ is proportional to ${\bf P}_e$, which is equivalent to the residual photocurrent in the PC model. 

In Figs. 3(a)-(d), we present $\boldsymbol{\hat{E}}_{\text{THz},i}$ and $|\boldsymbol{E}_{\text{THz},i}|$ of highest-weight trajectories with respect to $\varepsilon$ and ionization instants $t_0$ for two selected $\phi = 0,\pi/2$.
Figs. 3(e)-(h) analyze the TWGs from the trajectory ensemble, where the calculated $\boldsymbol{\hat{E}}_{\text{THz}}$, ${\bf \hat{P}}_e$ and measured THz polarization $\theta_\text{THz}$ are depicted in direct comparison with PMDs at $\varepsilon=0,1$. The corresponding simulation results of $\boldsymbol{\hat{E}}_{\text{THz},i}$, $|\boldsymbol{E}_{\text{THz},i}|$, $\boldsymbol{\hat{E}}_{\text{THz}}$, ${\bf \hat{P}}_e$ and PMDs without Coulomb potential are shown in Figs. 3 of \emph{Supplement} for comparison.

\begin{figure}[h]
  \includegraphics[width=0.8\textwidth]{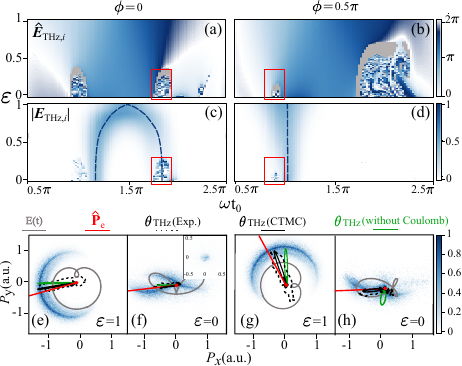}
  \caption{
  The trajectory analysis of THz emissions at various ellipticities $\varepsilon$ for two selected phase delay $\phi = 0,\pi/2$.
  (a)-(d) Contribution of the \emph{i}th individual trajectory to THz emission  $\boldsymbol{E}_{\text{THz},i}$ with respect to $\varepsilon$ and ionization instants $t_0$. Panel (a) and (c) show the polarization direction
  $\boldsymbol{\hat{E}}_{\text{THz},i}$ and the amplitude $|\boldsymbol{E}_{\text{THz},i}|$ at $\phi=0$. Panel (b) and (d) show $\boldsymbol{\hat{E}}_{\text{THz},i}$ and $|\boldsymbol{E}_{\text{THz},i}|$ at $\phi=\pi/2$.
  (e)-(h) THz emissions from the trajectory ensemble $\boldsymbol{E}_{\text{THz}}$. The PMDs, laser electric fields (grey bold lines), THz polarization $\boldsymbol{\hat{E}}_{\text{THz}}$ with (black solid lines) or without (green solid lines) Coulomb potential, asymmetry pointer of PMDs ${\bf \hat{P}}_e$ (red solid lines) and the experimental THz polarization $\theta_\text{THz}$ (dashed lines) are presented for comparison. The inset in Panel (f) depicts the PMD of selected trajectories in the red box in Panels (a) and (c).
  }\label{fig3}\label{fig:C-C-analysis} 
  
\end{figure}

For large ellipticities $\varepsilon > 0.4$, Figs. 3(a) and (b) illustrate that $\boldsymbol{\hat{E}}_{\text{THz},i}$ smoothly changes with respect to $t_0$. In Figs. 3(c) and (d), the maxima of $|\boldsymbol{E}_{\text{THz},i}|$ correspond to the peak values of the two-color fields (dashed lines), where the tunneling ionization rate reaches its maximum. 
These maxima represent the main tunneling temporal windows.
Figs. 3(e) and (g) for $\varepsilon$=$1$ exhibit that $\boldsymbol{\hat{E}}_{\text{THz}}$ (black solid lines) coincides with ${\bf \hat{P}}_e$ (red solid lines), 
as predicted in Eq. (1). 
The angular deviations observed between $\boldsymbol{\hat{E}}_{\text{THz}}$ with (black solid lines) and without (green solid lines) Coulomb potential result from the deflection of electron trajectory induced by Coulomb potential, i.e., soft collision between electron and parent ion. The angular deviation corresponds to the "streaking angle" in the "attoclock" of PMDs \cite{babushkin_all-optical_2022}.

As $\varepsilon$ decreases, chaos regions and grey regions emerge, highlighted in the red boxes in Figs. 3(a) and (b). 
The grey regions are explained by the scenario in which the parent core recaptures the free electron into the Rydberg state. 
The chaotic regions arise from the hard re-collision between the electron and parent core, resulting in the emission of $\boldsymbol{\hat{E}}_{\text{THz},i}$ occurring near-isotropically across all 4$\pi$ solid angles.
Trajectories within chaos regions that do not overlap with the maximum region of $|\boldsymbol{E}_{\text{THz},i}|$, as depicted in Figs. 3(b) and (d), contribute minimally to TWGs due to their low weight. 
However, when the chaos region overlaps with the right branch of tunneling windows, 
as red boxed
in Panel (a) and (c), the re-collision trajectories significantly influence the TWGs. 
The isotropic and isotropic distribution of $\boldsymbol{\hat{E}}_{\text{THz},i}$, as shown in the inset of Panel (f), leads to the counterbalancing of contributions from individual trajectories.
In Fig. 3(f), the angular deviations between $\boldsymbol{\hat{E}}_{\text{THz}}$ with and without Coulomb potential cannot be observed as in the case of $\varepsilon = 1$. 
It can be explained by the scenario that, although $\boldsymbol{\hat{E}}_{\text{THz},i}$ are deflected by Coulomb potential, yet the high-weight trajectories within the right branch of tunneling windows do not contribute to the TWGs,  the Coulomb potential thus is not effectively manifest in the TWGs.

When $\varepsilon$ changes from 1 to 0, the electron-core interaction transitions from soft collision to hard re-collision, manifested in the $\phi$-dependent TWGs. In hard re-collision, the random scattering breaks the homogeneous behavior of trajectory ensemble, diminishing the effectiveness of encoding dynamics and structural information in the TWGs. In contrast, during a soft collision, the Coulomb potential deflects the trajectory ensemble while maintaining its homogeneous behavior. In this scenario, the trajectory ensemble can be approximately represented by the highest-weight trajectory, providing a more straightforward basis for reconstructing electron dynamics.
This analysis can be further simplified
by an analytical solution obtained through perturbatively
evaluating the Coulomb-induced correction to the guiding center
trajectory, where the fast timescale laser-induced oscillation is averaged out
\cite{dubois_capturing_2018}. 
Refer to \emph{Supplement} Eqs. S5 and S6, and S15 for more details.

\section{Reconstruction of unknown potential by THz
radiation}\label{sec:reconstruct}

The all-optical reconstruction of the core potential of generating medium is conceptually straightforward. 
If a
sufficiently broad spectrum of the radiated field,
$\tilde{\boldsymbol{E}}_{\text{rad}} (\omega)$, can be acquired, the
reconstructed acceleration of photoelectron, $\boldsymbol{a} (t)
=\boldsymbol{E}_{\text{rad}} (t)$, in principle, allows tracing the local
potential, $\nabla V (\boldsymbol{r}) = -\boldsymbol{a}_C (\boldsymbol{r} (t)) = - [\boldsymbol{a}
(t) +\boldsymbol{a_0} (t)]$. 
Multiple trajectories under different  phase delay $\phi$, therefore, sketch the
contour of $\nabla V (\boldsymbol{r})$ as analogous to the mesh representation
of an object in an artistic wire sculpture (See \emph{Supplement} Sec.5 for details).

\begin{figure}[h]
  \includegraphics[width=0.9\textwidth]{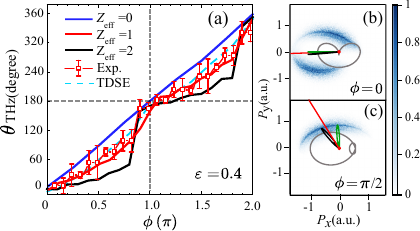}
  \caption{
   Reconstruction for the effective Coulomb potential with THz polarizations. (a) Dependence of the THz polarization direction $\theta_{\tmop{THz}} (\phi)$ on time delay $\phi$ with different effective charge $Z_{\tmop{eff}}$ at $\varepsilon$=0.4. (b) and (c) the same plots as Fig. \ref{fig3} (e)-(h), but for $\varepsilon$=$0.4$, $Z_{\tmop{eff}} $=1
   at $\phi$=0 and $\pi$/2 respectively.} \label{fig4}\label{fig:potential-reconstruction}

\end{figure}

The acceleration $\boldsymbol{a}_C$ induced by Coulomb potential becomes significant only when 
the electron-nucleus distance
$\boldsymbol{r} (t)$ is very small. Consequently, Coulomb potential seriously modifies the TWG during the initial phase when the electron has just departed from the core after ionization. The TWG modification within a narrow temporal windows corresponds to the high-frequency component of THz emissions. 
Theoretical testing suggests that the accurate reconstruction of the entire profile of the Coulomb potential would necessitate a broad-spectrum THz coverage up to 280 THz. 
However, despite recent significant advancements in generating and detecting broadband THz emissions in gas plasma, the capabilities of TWG generation and detection remain restricted to 100 THz.

The practical limitation in our setup allows for reliable measurement only from 0.1 to 3 THz.
Fortunately, the TWG is determined by
the slow timescale dynamics that are highly sensitive to the initial stage of
the photoelectron motion. As the interaction with the parent ion can
dramatically alter the photoelectron trajectory when the electron roams around
the core within a short time after the tunneling ionization, 
it is hence
still possible to extract partial information by exploiting the TWG.

It can be shown by an example to retrieve a key parameter of the effective potential. We assume a Coulomb potential $V (\boldsymbol{r}) = - \frac{Z_{\tmop{eff}}}{| \boldsymbol{r} |}$ with $Z_{\tmop{eff}}$ the effective charge, which reflects the strength of the Coulomb potential, to be determined. 

The polarization direction of TWG as a functon of phase delay $\phi$, i.e. $\theta_\text{THz}(\phi)$, can be experimentally measured by scanning $\phi$. The absolute $\phi$ can be determined by comparing the measurement and theories, including CTMC and direct solution of the time-dependent Schr\"odinger equation (TDSE). As shown in Fig. 4(a), $\theta_\text{THz} = 0^{\circ},180^{\circ}$ at $\phi = 0^{\circ},180^{\circ}$ remains identical regardless of $Z_{\mathrm{eff}}$, establishing a criterion for determining the absolute $\phi$ (see \emph{Supplement} S.7 for detailed calibration).

As shown in Fig. 4(a), $\theta_\text{THz}(\phi)$ possesses a high correlation with $Z_{\tmop{eff}}$, allowing for the determination of $Z_{\tmop{eff}}$ by comparing the experimentally obtained $\theta_\text{THz}(\phi)$ with the simulation one.
Figs. 4(b) and (c) presents the same plots as Fig.3 (e)-(h) for $\varepsilon=0.4$, $\phi=0,\pi/2$ as a further inspection of the comparison of TWG polarization and angular streaking in PMDs. In Fig. 4, both $\theta_\text{THz}$ and $\hat{\bf {P}}_e$ are rotated as a function of $\phi$, establishing a connection between our measurement and the "phase-of-phase (POP) attoclock" experiment\cite{Liu_terahertz_2021}. $\theta_\text{THz}(\phi)$ in our measurement and the emitting angle in "POP attoclock" show the similar evolution with respect to $\phi$, validating the reconstruction methodology based on TWGs. The congruence implies the THz polarization measurements could serve as an alternative to the "POP attoclock", providing a potential avenue for extracting tunneling ionization dynamics in future studies.

\section{Conclusion}

In this work, we found that the dependence of THz yields on the ellipticity and interpulse phase delay of bi-chromatic laser cannot be explained by the PC model due to the absence of the photoelectron-core interaction. The inclusion of Coulomb potential in the CTMC model not only reproduces the experimental results but also establishes the connection between THz radiation and photoelectron motion. 
Compared to the re-collision scenario at low ellipticity,
the structure information is more efficiently encoded in the motion of trajectory ensemble after a soft collision at intermediate ellipticity. 
Therefore, with the support of CTMC and TDSE simulation, we introduce a reconstruction methodology for extracting local potential by measuring THz polarizations with respect to the two-color phase delay. The THz polarization is equivalent to the asymmetry pointer of PMDs, connecting our measurement and "attoclock" of PMDs. 
In contrast to the angular offset in the conventional "attoclock", our experiment allows for a  precise and easy-to-implement determination of THz polarization.
Furthermore, TWGs emitted from condensed-phase media provide an opportunity to extract electron motion and structure in the bulk solid or liquid targets.


\begin{acknowledgement}

This work was supported by the National Key Research and Development Program of China (No. 2022YFA1604302) and the National Natural Science Foundation of China (No. 12334011, No. 12174284, and No.12374262). We acknowledge the computational resource of the HPC platform in ShanghaiTech University. 

\end{acknowledgement}


\bibliography{apssamp.bib}

\providecommand{\latin}[1]{#1}
\makeatletter
\providecommand{\doi}
  {\begingroup\let\do\@makeother\dospecials
  \catcode`\{=1 \catcode`\}=2 \doi@aux}
\providecommand{\doi@aux}[1]{\endgroup\texttt{#1}}
\makeatother
\providecommand*\mcitethebibliography{\thebibliography}
\csname @ifundefined\endcsname{endmcitethebibliography}
  {\let\endmcitethebibliography\endthebibliography}{}
\begin{mcitethebibliography}{31}
\providecommand*\natexlab[1]{#1}
\providecommand*\mciteSetBstSublistMode[1]{}
\providecommand*\mciteSetBstMaxWidthForm[2]{}
\providecommand*\mciteBstWouldAddEndPuncttrue
  {\def\EndOfBibitem{\unskip.}}
\providecommand*\mciteBstWouldAddEndPunctfalse
  {\let\EndOfBibitem\relax}
\providecommand*\mciteSetBstMidEndSepPunct[3]{}
\providecommand*\mciteSetBstSublistLabelBeginEnd[3]{}
\providecommand*\EndOfBibitem{}
\mciteSetBstSublistMode{f}
\mciteSetBstMaxWidthForm{subitem}{(\alph{mcitesubitemcount})}
\mciteSetBstSublistLabelBeginEnd
  {\mcitemaxwidthsubitemform\space}
  {\relax}
  {\relax}

\bibitem[Itatani \latin{et~al.}(2004)Itatani, Levesque, Zeidler, Niikura,
  P\'{e}pin, Kieffer, Corkum, and Villeneuve]{itatani_tomographic_2004}
Itatani,~J.; Levesque,~J.; Zeidler,~D.; Niikura,~H.; P\'{e}pin,~H.;
  Kieffer,~J.~C.; Corkum,~P.~B.; Villeneuve,~D.~M. Tomographic imaging of
  molecular orbitals. \emph{Nature} \textbf{2004}, \emph{432}, 867--871\relax
\mciteBstWouldAddEndPuncttrue
\mciteSetBstMidEndSepPunct{\mcitedefaultmidpunct}
{\mcitedefaultendpunct}{\mcitedefaultseppunct}\relax
\EndOfBibitem
\bibitem[Vozzi \latin{et~al.}(2011)Vozzi, Negro, Calegari, Sansone, Nisoli,
  De~Silvestri, and Stagira]{vozzi_generalized_2011}
Vozzi,~C.; Negro,~M.; Calegari,~F.; Sansone,~G.; Nisoli,~M.; De~Silvestri,~S.;
  Stagira,~S. Generalized molecular orbital tomography. \emph{Nat. Phys.}
  \textbf{2011}, \emph{7}, 822--826\relax
\mciteBstWouldAddEndPuncttrue
\mciteSetBstMidEndSepPunct{\mcitedefaultmidpunct}
{\mcitedefaultendpunct}{\mcitedefaultseppunct}\relax
\EndOfBibitem
\bibitem[Shafir \latin{et~al.}(2012)Shafir, Soifer, Bruner, Dagan, Mairesse,
  Patchkovskii, Ivanov, Smirnova, and Dudovich]{shafir_resolving_2012}
Shafir,~D.; Soifer,~H.; Bruner,~B.~D.; Dagan,~M.; Mairesse,~Y.;
  Patchkovskii,~S.; Ivanov,~M.~Y.; Smirnova,~O.; Dudovich,~N. Resolving the
  time when an electron exits a tunnelling barrier. \emph{Nature}
  \textbf{2012}, \emph{485}, 343--346\relax
\mciteBstWouldAddEndPuncttrue
\mciteSetBstMidEndSepPunct{\mcitedefaultmidpunct}
{\mcitedefaultendpunct}{\mcitedefaultseppunct}\relax
\EndOfBibitem
\bibitem[Pedatzur \latin{et~al.}(2015)Pedatzur, Orenstein, Serbinenko, Soifer,
  Bruner, Uzan, Brambila, Harvey, Torlina, Morales, Smirnova, and
  Dudovich]{pedatzur_attosecond_2015}
Pedatzur,~O.; Orenstein,~G.; Serbinenko,~V.; Soifer,~H.; Bruner,~B.~D.;
  Uzan,~A.~J.; Brambila,~D.~S.; Harvey,~A.~G.; Torlina,~L.; Morales,~F.;
  Smirnova,~O.; Dudovich,~N. Attosecond tunnelling interferometry. \emph{Nat.
  Phys.} \textbf{2015}, \emph{11}, 815--U184\relax
\mciteBstWouldAddEndPuncttrue
\mciteSetBstMidEndSepPunct{\mcitedefaultmidpunct}
{\mcitedefaultendpunct}{\mcitedefaultseppunct}\relax
\EndOfBibitem
\bibitem[Lein(2005)]{lein_attosecond_2005}
Lein,~M. Attosecond {Probing} of {Vibrational} {Dynamics} with
  {High}-{Harmonic} {Generation}. \emph{Phys. Rev. Lett.} \textbf{2005},
  \emph{94}, 053004\relax
\mciteBstWouldAddEndPuncttrue
\mciteSetBstMidEndSepPunct{\mcitedefaultmidpunct}
{\mcitedefaultendpunct}{\mcitedefaultseppunct}\relax
\EndOfBibitem
\bibitem[Baker \latin{et~al.}(2006)Baker, Robinson, Haworth, Teng, Smith,
  Chiril\u{a}, Lein, Tisch, and Marangos]{baker_probing_2006}
Baker,~S.; Robinson,~J.~S.; Haworth,~C.~A.; Teng,~H.; Smith,~R.~A.;
  Chiril\u{a},~C.~C.; Lein,~M.; Tisch,~J. W.~G.; Marangos,~J.~P. Probing
  {Proton} {Dynamics} in {Molecules} on an {Attosecond} {Time} {Scale}.
  \emph{Science} \textbf{2006}, \emph{312}, 424\relax
\mciteBstWouldAddEndPuncttrue
\mciteSetBstMidEndSepPunct{\mcitedefaultmidpunct}
{\mcitedefaultendpunct}{\mcitedefaultseppunct}\relax
\EndOfBibitem
\bibitem[Bian and Bandrauk(2014)Bian, and Bandrauk]{bian_probing_2014}
Bian,~X.-B.; Bandrauk,~A.~D. Probing {Nuclear} {Motion} by {Frequency}
  {Modulation} of {Molecular} {High}-{Order} {Harmonic} {Generation}.
  \emph{Phys. Rev. Lett.} \textbf{2014}, \emph{113}, 193901\relax
\mciteBstWouldAddEndPuncttrue
\mciteSetBstMidEndSepPunct{\mcitedefaultmidpunct}
{\mcitedefaultendpunct}{\mcitedefaultseppunct}\relax
\EndOfBibitem
\bibitem[Cook and Hochstrasser(2000)Cook, and Hochstrasser]{cook_intense_2000}
Cook,~D.~J.; Hochstrasser,~R.~M. Intense terahertz pulses by four-wave
  rectification in air. \emph{Opt. Lett} \textbf{2000}, \emph{25},
  1210--1212\relax
\mciteBstWouldAddEndPuncttrue
\mciteSetBstMidEndSepPunct{\mcitedefaultmidpunct}
{\mcitedefaultendpunct}{\mcitedefaultseppunct}\relax
\EndOfBibitem
\bibitem[Brunel(1990)]{Brunel_Harmonic_1990}
Brunel,~F. Harmonic generation due to plasma effects in a gas undergoing
  multiphoton ionization in the high-intensity limit. \emph{J. Opt. Soc. Am. B}
  \textbf{1990}, \emph{7}, 521--526\relax
\mciteBstWouldAddEndPuncttrue
\mciteSetBstMidEndSepPunct{\mcitedefaultmidpunct}
{\mcitedefaultendpunct}{\mcitedefaultseppunct}\relax
\EndOfBibitem
\bibitem[Huang \latin{et~al.}(2015)Huang, Meng, Wang, L\"u, Zhang, Chen, Zhao,
  Yuan, and Zhao]{huang_joint_2015}
Huang,~Y.; Meng,~C.; Wang,~X.; L\"u,~Z.; Zhang,~D.; Chen,~W.; Zhao,~J.;
  Yuan,~J.; Zhao,~Z. Joint Measurements of Terahertz Wave Generation and
  High-Harmonic Generation from Aligned Nitrogen Molecules Reveal
  Angle-Resolved Molecular Structures. \emph{Phys. Rev. Lett.} \textbf{2015},
  \emph{115}, 123002\relax
\mciteBstWouldAddEndPuncttrue
\mciteSetBstMidEndSepPunct{\mcitedefaultmidpunct}
{\mcitedefaultendpunct}{\mcitedefaultseppunct}\relax
\EndOfBibitem
\bibitem[Babushkin \latin{et~al.}(2022)Babushkin, Gal\'{a}n, de~Andrade,
  Husakou, Morales, Kretschmar, Nagy, Vai\v{c}aitis, Shi, Zuber, Berg\'{e},
  Skupin, Nikolaeva, Panov, Shipilo, Kosareva, Pfeiffer, Demircan, Vrakking,
  Morgner, and Ivanov]{babushkin_all-optical_2022}
Babushkin,~I. \latin{et~al.}  All-optical attoclock for imaging tunnelling
  wavepackets. \emph{Nat. Phys.} \textbf{2022}, \emph{18}, 417--422\relax
\mciteBstWouldAddEndPuncttrue
\mciteSetBstMidEndSepPunct{\mcitedefaultmidpunct}
{\mcitedefaultendpunct}{\mcitedefaultseppunct}\relax
\EndOfBibitem
\bibitem[Eckle \latin{et~al.}(2008)Eckle, Smolarski, Schlup, Biegert, Staudte,
  Schoeffler, Muller, Doerner, and Keller]{Eckle_attosecond_2008a}
Eckle,~P.; Smolarski,~M.; Schlup,~P.; Biegert,~J.; Staudte,~A.; Schoeffler,~M.;
  Muller,~H.~G.; Doerner,~R.; Keller,~U. Attosecond angular streaking.
  \emph{Nat. Phys.} \textbf{2008}, \emph{4}, 565--570\relax
\mciteBstWouldAddEndPuncttrue
\mciteSetBstMidEndSepPunct{\mcitedefaultmidpunct}
{\mcitedefaultendpunct}{\mcitedefaultseppunct}\relax
\EndOfBibitem
\bibitem[Eckle \latin{et~al.}(2008)Eckle, Pfeiffer, Cirelli, Staudte, Doerner,
  Muller, Buettiker, and Keller]{Eckle_attosecond_2008b}
Eckle,~P.; Pfeiffer,~A.~N.; Cirelli,~C.; Staudte,~A.; Doerner,~R.;
  Muller,~H.~G.; Buettiker,~M.; Keller,~U. Attosecond Ionization and Tunneling
  Delay Time Measurements in Helium. \emph{Science} \textbf{2008}, \emph{322},
  1525--1529\relax
\mciteBstWouldAddEndPuncttrue
\mciteSetBstMidEndSepPunct{\mcitedefaultmidpunct}
{\mcitedefaultendpunct}{\mcitedefaultseppunct}\relax
\EndOfBibitem
\bibitem[Camus \latin{et~al.}(2017)Camus, Yakaboylu, Fechner, Klaiber, Laux,
  Mi, Hatsagortsyan, Pfeifer, Keitel, and Moshammer]{camus_experimental_2017}
Camus,~N.; Yakaboylu,~E.; Fechner,~L.; Klaiber,~M.; Laux,~M.; Mi,~Y.;
  Hatsagortsyan,~K.~Z.; Pfeifer,~T.; Keitel,~C.~H.; Moshammer,~R. Experimental
  Evidence for Quantum Tunneling Time. \emph{Phys. Rev. Lett.} \textbf{2017},
  \emph{119}, 023201\relax
\mciteBstWouldAddEndPuncttrue
\mciteSetBstMidEndSepPunct{\mcitedefaultmidpunct}
{\mcitedefaultendpunct}{\mcitedefaultseppunct}\relax
\EndOfBibitem
\bibitem[Sainadh \latin{et~al.}(2019)Sainadh, Xu, Wang, Atia-Tul-Noor, Wallace,
  Douguet, Bray, Ivanov, Bartschat, Kheifets, Sang, and
  Litvinyuk]{Sainadh_attosecond_2019}
Sainadh,~U.~S.; Xu,~H.; Wang,~X.; Atia-Tul-Noor,~A.; Wallace,~W.~C.;
  Douguet,~N.; Bray,~A.; Ivanov,~I.; Bartschat,~K.; Kheifets,~A.; Sang,~R.~T.;
  Litvinyuk,~I.~V. Attosecond angular streaking and tunnelling time in atomic
  hydrogen (vol 568, pg 75, 2019). \emph{Nature} \textbf{2019}, \emph{569},
  E5\relax
\mciteBstWouldAddEndPuncttrue
\mciteSetBstMidEndSepPunct{\mcitedefaultmidpunct}
{\mcitedefaultendpunct}{\mcitedefaultseppunct}\relax
\EndOfBibitem
\bibitem[Kim \latin{et~al.}(2007)Kim, Glownia, Taylor, and
  Rodriguez]{kim_terahertz_2007}
Kim,~K.~Y.; Glownia,~J.~H.; Taylor,~A.~J.; Rodriguez,~G. Terahertz emission
  from ultrafast ionizing air in symmetry-broken laser fields. \emph{Opt.
  Express} \textbf{2007}, \emph{15}, 4577--4584\relax
\mciteBstWouldAddEndPuncttrue
\mciteSetBstMidEndSepPunct{\mcitedefaultmidpunct}
{\mcitedefaultendpunct}{\mcitedefaultseppunct}\relax
\EndOfBibitem
\bibitem[Zhou \latin{et~al.}(2009)Zhou, Zhang, Zhao, and
  Yuan]{zhou_terahertz_2009}
Zhou,~Z.; Zhang,~D.; Zhao,~Z.; Yuan,~J. Terahertz emission of atoms driven by
  ultrashort laser pulses. \emph{Phys. Rev. A} \textbf{2009}, \emph{79},
  063413\relax
\mciteBstWouldAddEndPuncttrue
\mciteSetBstMidEndSepPunct{\mcitedefaultmidpunct}
{\mcitedefaultendpunct}{\mcitedefaultseppunct}\relax
\EndOfBibitem
\bibitem[Zhang \latin{et~al.}(2020)Zhang, Zhang, Wang, Yan, and
  Jiang]{zhang_continuum_2020}
Zhang,~K.; Zhang,~Y.; Wang,~X.; Yan,~T.-M.; Jiang,~Y.~H. Continuum electron
  giving birth to terahertz emission. \emph{Photonics Res.} \textbf{2020},
  \emph{8}, 760--767\relax
\mciteBstWouldAddEndPuncttrue
\mciteSetBstMidEndSepPunct{\mcitedefaultmidpunct}
{\mcitedefaultendpunct}{\mcitedefaultseppunct}\relax
\EndOfBibitem
\bibitem[Azoury \latin{et~al.}(2019)Azoury, Kneller, Rozen, Bruner, Clergerie,
  Mairesse, Fabre, Pons, Dudovich, and Kruger]{Azoury_electronic_2019}
Azoury,~D.; Kneller,~O.; Rozen,~S.; Bruner,~B.~D.; Clergerie,~A.; Mairesse,~Y.;
  Fabre,~B.; Pons,~B.; Dudovich,~N.; Kruger,~M. Electronic wavefunctions probed
  by all-optical attosecond interferometry. \emph{Nat. Photonics}
  \textbf{2019}, \emph{13}, 54\relax
\mciteBstWouldAddEndPuncttrue
\mciteSetBstMidEndSepPunct{\mcitedefaultmidpunct}
{\mcitedefaultendpunct}{\mcitedefaultseppunct}\relax
\EndOfBibitem
\bibitem[Boutu \latin{et~al.}(2008)Boutu, Haessler, Merdji, Breger, Waters,
  Stankiewicz, Frasinski, Taieb, Caillat, Maquet, Monchicourt, and
  Carre]{Boutu_2008}
Boutu,~W.; Haessler,~S.; Merdji,~H.; Breger,~P.; Waters,~G.; Stankiewicz,~M.;
  Frasinski,~L.; Taieb,~R.; Caillat,~J.; Maquet,~A.; Monchicourt,~P.; Carre,~B.
  Coherent control of attosecond emission from aligned molecules. \emph{Nat.
  Phys.} \textbf{2008}, \emph{4}, 545--549\relax
\mciteBstWouldAddEndPuncttrue
\mciteSetBstMidEndSepPunct{\mcitedefaultmidpunct}
{\mcitedefaultendpunct}{\mcitedefaultseppunct}\relax
\EndOfBibitem
\bibitem[Zhou \latin{et~al.}(2008)Zhou, Lock, Li, Wagner, Murnane, and
  Kapteyn]{Zhou_2008}
Zhou,~X.; Lock,~R.; Li,~W.; Wagner,~N.; Murnane,~M.~M.; Kapteyn,~H.~C.
  Molecular Recollision Interferometry in High Harmonic Generation. \emph{Phys.
  Rev. Lett.} \textbf{2008}, \emph{100}, 073902\relax
\mciteBstWouldAddEndPuncttrue
\mciteSetBstMidEndSepPunct{\mcitedefaultmidpunct}
{\mcitedefaultendpunct}{\mcitedefaultseppunct}\relax
\EndOfBibitem
\bibitem[Kim \latin{et~al.}(2008)Kim, Taylor, Glownia, and
  Rodriguez]{kim_coherent_2008}
Kim,~K.~Y.; Taylor,~A.~J.; Glownia,~J.~H.; Rodriguez,~G. Coherent control of
  terahertz supercontinuum generation in ultrafast laser-gas interactions.
  \emph{Nat. Photonics} \textbf{2008}, \emph{2}, 605--609\relax
\mciteBstWouldAddEndPuncttrue
\mciteSetBstMidEndSepPunct{\mcitedefaultmidpunct}
{\mcitedefaultendpunct}{\mcitedefaultseppunct}\relax
\EndOfBibitem
\bibitem[Babushkin \latin{et~al.}(2011)Babushkin, Skupin, Husakou, K\"{o}hler,
  Cabrera-Granado, Berg\'{e}, and Herrmann]{Babushkin_2011}
Babushkin,~I.; Skupin,~S.; Husakou,~A.; K\"{o}hler,~C.; Cabrera-Granado,~E.;
  Berg\'{e},~L.; Herrmann,~J. Tailoring terahertz radiation by controlling
  tunnel photoionization events in gases. \emph{New J. Phys} \textbf{2011},
  \emph{13}, 123029\relax
\mciteBstWouldAddEndPuncttrue
\mciteSetBstMidEndSepPunct{\mcitedefaultmidpunct}
{\mcitedefaultendpunct}{\mcitedefaultseppunct}\relax
\EndOfBibitem
\bibitem[Liu \latin{et~al.}(2014)Liu, Chen, Zhang, Zhao, Wu, Yuan, and
  Zhao]{liu_trajectory-based_2014}
Liu,~J.; Chen,~W.; Zhang,~B.; Zhao,~J.; Wu,~J.; Yuan,~J.; Zhao,~Z.
  Trajectory-based analysis of low-energy electrons and photocurrents generated
  in strong-field ionization. \emph{Phys. Rev. A} \textbf{2014}, \emph{90},
  063420\relax
\mciteBstWouldAddEndPuncttrue
\mciteSetBstMidEndSepPunct{\mcitedefaultmidpunct}
{\mcitedefaultendpunct}{\mcitedefaultseppunct}\relax
\EndOfBibitem
\bibitem[Zhang \latin{et~al.}(2012)Zhang, L\"{u}, Meng, Du, Zhou, Zhao, and
  Yuan]{zhang_synchronizing_2012}
Zhang,~D.; L\"{u},~Z.; Meng,~C.; Du,~X.; Zhou,~Z.; Zhao,~Z.; Yuan,~J.
  Synchronizing {Terahertz} {Wave} {Generation} with {Attosecond} {Bursts}.
  \emph{Phys. Rev. Lett.} \textbf{2012}, \emph{109}, 243002\relax
\mciteBstWouldAddEndPuncttrue
\mciteSetBstMidEndSepPunct{\mcitedefaultmidpunct}
{\mcitedefaultendpunct}{\mcitedefaultseppunct}\relax
\EndOfBibitem
\bibitem[Gao \latin{et~al.}(2023)Gao, Zhang, Zhang, Gan, Yan, and
  Jiang]{gao_2023_coulomb}
Gao,~Y.; Zhang,~Y.; Zhang,~K.; Gan,~Z.; Yan,~T.-M.; Jiang,~Y. Coulomb potential
  determining terahertz polarization in a two-color laser field. \emph{Opt.
  Lett.} \textbf{2023}, \emph{48}, 2575--2578\relax
\mciteBstWouldAddEndPuncttrue
\mciteSetBstMidEndSepPunct{\mcitedefaultmidpunct}
{\mcitedefaultendpunct}{\mcitedefaultseppunct}\relax
\EndOfBibitem
\bibitem[Zhang \latin{et~al.}(2021)Zhang, Zhang, Yan, and
  Jiang]{zhang_electron_2021}
Zhang,~Y.; Zhang,~K.; Yan,~T.-M.; Jiang,~Y. Electron trajectory backanalysis
  for spectral profile in two-color terahertz generation. \emph{J. Phys. B: At.
  Mol. Opt. Phys.} \textbf{2021}, \emph{54}, 195401\relax
\mciteBstWouldAddEndPuncttrue
\mciteSetBstMidEndSepPunct{\mcitedefaultmidpunct}
{\mcitedefaultendpunct}{\mcitedefaultseppunct}\relax
\EndOfBibitem
\bibitem[Fan \latin{et~al.}(2023)Fan, Gao, Yan, Jiang, and Zhang]{Fan_2023}
Fan,~X.; Gao,~Y.; Yan,~T.-M.; Jiang,~Y.; Zhang,~Y. Trajectory analysis for
  low-order harmonic generation in two-color strong laser fields. \emph{Opt.
  Express} \textbf{2023}, \emph{31}, 86--94\relax
\mciteBstWouldAddEndPuncttrue
\mciteSetBstMidEndSepPunct{\mcitedefaultmidpunct}
{\mcitedefaultendpunct}{\mcitedefaultseppunct}\relax
\EndOfBibitem
\bibitem[Dubois \latin{et~al.}(2018)Dubois, Berman, Chandre, and
  Uzer]{dubois_capturing_2018}
Dubois,~J.; Berman,~S.~A.; Chandre,~C.; Uzer,~T. Capturing Photoelectron Motion
  with Guiding Centers. \emph{Phys. Rev. Lett.} \textbf{2018}, \emph{121},
  113202\relax
\mciteBstWouldAddEndPuncttrue
\mciteSetBstMidEndSepPunct{\mcitedefaultmidpunct}
{\mcitedefaultendpunct}{\mcitedefaultseppunct}\relax
\EndOfBibitem
\bibitem[Han \latin{et~al.}(2021)Han, Ge, Wang, Zhenning, Fang, Ma, Yu, Deng,
  W\"orner, Gong, and Liu]{Liu_terahertz_2021}
Han,~M.; Ge,~P.; Wang,~J.; Zhenning,~G.; Fang,~Y.; Ma,~X.; Yu,~Y.; Deng,~Y.;
  W\"orner,~H.~J.; Gong,~Q.; Liu,~Y. Complete characterization of
  sub-Coulomb-barrier tunnelling with phase-of-phase attoclock. \emph{Nat.
  Photonics} \textbf{2021}, \emph{15}, 765--771\relax
\mciteBstWouldAddEndPuncttrue
\mciteSetBstMidEndSepPunct{\mcitedefaultmidpunct}
{\mcitedefaultendpunct}{\mcitedefaultseppunct}\relax
\EndOfBibitem
\end{mcitethebibliography}

\end{document}